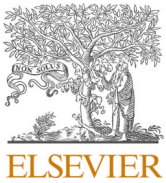
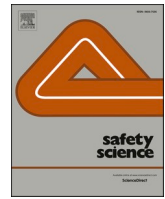

# Development of an end-to-end hardware and software pipeline for affordable and feasible ergonomics assessment in the automotive industry

J. González-Alonso [a,*], C. Simón-Martínez [b], M. Antón-Rodríguez [a], D. González-Ortega [a], F.J. Díaz-Pernas [a], M. Martínez-Zarzuela [a]

[a] *Department of Signal Theory, Communications and Telematics Engineering, Telecommunications Engineering School, University of Valladolid, 47011 Valladolid, Spain*
[b] *University of Applied Sciences and Arts Western Switzerland (HES-SO) Valais-Wallis, Sierre Institute of Informatics, Rue de Technopole 3, 3965 Sierre, Switzerland*

A B S T R A C T

An end-to-end hardware-software pipeline is introduced to automatize ergonomics assessment in industrial workplaces. The proposed modular solution can interoperate with commercial systems throughout the ergonomics assessment phases involved in the process. The pipeline includes custom-designed Inertial Measurement Unit (IMU) sensors, two real-time worker movement acquisition tools, inverse kinematics processing and Rapid Upper Limb Assessment (RULA) report generation. It is based on free tools such as Unity3D and OpenSim to avoid the problems derived from using proprietary technologies, such as security decisions being made under "black box" conditions. Experiments were conducted in an automotive factory in a workplace with WMSDs risk among workers. The proposed solution obtained comparable results to a gold standard solution, reaching measured joint angles a 0.95 cross-correlation and a Root Mean Square Error (RMSE) lower than 10 for elbows and 12 for shoulders between both systems. In addition, the global RULA score difference is lower than 5 % between both systems. This work provides a low-cost solution for WMSDs risk assessment in the workplace to reduce musculoskeletal disorders and associated sick leave in industry, impacting the health of workers in the long term. Our study can ease further research and popularize the use of wearable systems for ergonomics analysis allowing these workplace prevention systems to reach different industrial environments.

## 1. Introduction

The most common occupational diseases in the European Union are related to musculoskeletal disorders, which affect workers in all sectors and professions (Work-related musculoskeletal disorders: prevalence, costs and demographics in the EU, 2022). Ergonomic stressors such as biomechanical constraints, awkward postures, and repetitive movements are closely related to soft-tissue injuries among workers when they perform physical tasks (Cole et al., 2005). In recent years, there has been increasing interest in measuring human motion and Work-related Musculoskeletal Disorders (WMSDs) risk to improve posture and workers' safety in industrial workplaces (Menolotto et al., 2020). The ergonomics assessment becomes critical in industry, as workers sick leave and occupational accidents generate considerable associated costs. The primary cost contributor is its impact on production, surpassing expenses related to sickness absence and medical assistance (Rosado et al., 2023). In the European Union, the profound economic impact of WMSDs is exemplified by Germany, which experienced a staggering production loss of EUR 17.2 billion in 2016, calculated based on labor costs, along with a subsequent reduction in gross value added by EUR 30.4 billion (Work-related musculoskeletal disorders: prevalence, costs and demographics in the EU, 2022). Similarly, a parallel examination of costs in Great Britain in 2013/14 arising from new WMSDs cases reveals an estimated annual social cost of approximately £2.3 billion (Health and safety statistics, 2021). These figures highlight the significant contribution of the economic burden of WMSDs in the European countries.

Traditionally, postural safety testing at work is performed by ergonomists using direct observation and video recordings (Maldonado et al., 2015). Over time, various methods have been employed to assess the risk of WMSDs in workstations on production lines. Most of these methods are based on the prolonged maintenance of certain incorrect postures, which are determined by joint angles. Workstations are eventually modified according to the results of these assessment methods. Among them, a commonly used method is the Rapid Upper Limb Assessment (RULA) (McAtamney et al., 1993), which evaluates the posture and movement of upper limb joints during a specific task. The RULA method is widely accepted as reliable and valid for identifying and assessing WMSDs risk in the workplace. However, not only is the ergonomics assessment procedure relatively subjective, but also the ergonomist is often only required to assess a workstation when injuries have already appeared, since the analysis reveals the adaptation needs of the workplace a posteriori. In this case, workplace adaptation cannot be introduced preventively during the design phase of the workplace or the first few times it is used. For this reason, human body tracking systems that can assist ergonomists in more frequent assessments of workplaces with quantitative data are highly desirable. Embracing such technology can enhance the effectiveness of ergonomic assessments and contribute to healthier work environments (Salisu et al., 2023).

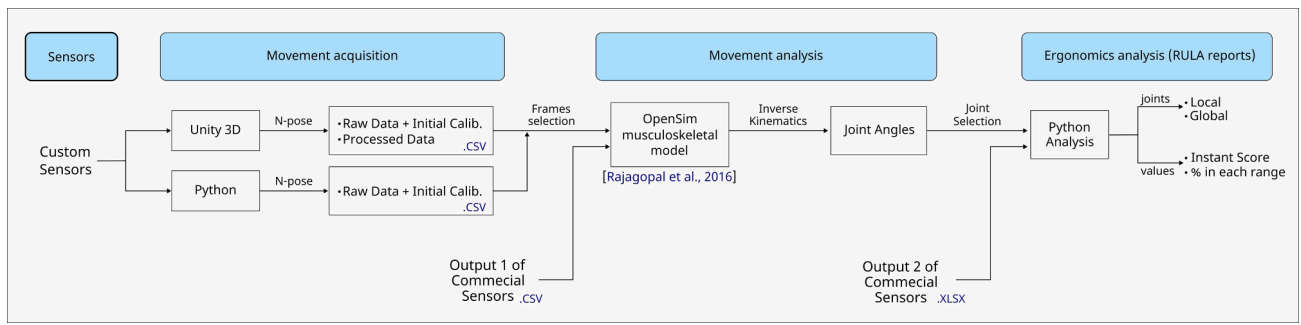

**Fig. 1.** Overall diagram of the presented pipeline.

Motion capture in industrial environments has become valuable for ergonomics because it enables the accurate and objective measurement of human movement and posture. This capability can be utilized to assess the risk of musculoskeletal disorders in the workplace and provide valuable insights for evaluating the effectiveness of ergonomics interventions over time (Menolotto et al., 2020; Vijayakumar and Choi, 2022; Ranavolo et al., 2018). Previous studies have introduced computer methods based on different hardware-software approaches (Vignais et al., 2013; Vignais et al., 2017; Caputo et al., 2019; Maurice et al., 2019; Greco et al., 2020; Huang et al., 2020; Colim et al., 2021; Panariello et al., 2022) and recent studies have confirmed the feasibility of using inertial sensor-based solutions for body motion capture (Sers et al., 2020; Carnevale et al., 2019; Taylor et al., 2017; Poitras et al., 2019) in real-life conditions. These solutions offer certain advantages over those based on computer vision, including the possibility to perform measurements in industrial workplaces without the need for a dedicated space. For this reason, motion capture systems using Inertial Measurement Units (IMU) have become increasingly important to study human movement (López-Nava and Muñoz-Meléndez, 2016). Some commercial systems can be used as ergonomics tools to capture and analyze postural behavior. For example, Movella Awinda (Paulich et al., 2018) provides a complex capture and analysis suite called MVN Analyze Pro, which gives useful information on joint angles and ergonomics, and it has been validated against multi-camera body tracking systems using computer vision (Robert-Lachaine et al., 2017).

Although systems using video and positional markers (Nexus Vicon, Optitrak) are still considered the gold standard in the field of clinical research for postural studies, they have some disadvantages that prevent their use for WMSDs assessment in outside-of-the-lab settings. Among other limitations, we can find their complexity of use, their high cost, and large dimensions, as well as the need for a large and dedicated space; and thus, not adjusting to the reality of industrial workstations (Salisu et al., 2023). As a result, the video approach has been mainly applied to a couple of industries such as construction, robot manufacturing, and other generic applications (Menolotto et al., 2020) and is not a widespread practice in the rest of industries. Moreover, these vision-based systems also preclude in situ capture as they require dedicated space or modifications to the workplace to prevent disruptions to the standard movements performed by workers (Salisu et al., 2023; Ranavolo et al., 2018).

Commercial IMU-based solutions such as Movella Awinda (Huang et al., 2020; Colim et al., 2021) are becoming the de facto gold standard for WMSDs assessment in the workplace. Nevertheless, these commercial solutions come with subscription-oriented and stand-alone programs for analysis, increasing their cost and decreasing the transparency of the analysis behind the pay-wall. One of their main drawbacks is that they are closed solutions and do not allow ergonomists and researchers to control simulation parameters. Differently, custom-developed systems for IMU-based body motion capture can be used to obtain data measurements of the orientation of human body parts in real time and accurate joint angles using inverse kinematics. A growing number of studies have found that low-cost custom IMU-based solutions can satisfy human body motion estimation requirements (Caputo et al., 2019; Greco et al., 2020; López-Nava and Muñoz-Meléndez, 2016). Still, some of them are only applicable when there are no ferromagnetic materials nearby (Vignais et al., 2013; Álvarez et al., 2016), making them unsuitable for certain working conditions in the automotive industry; or they focus only on providing a hardware solution for the motion capture process without covering the subsequent ergonomics analysis software (He et al., 2022; Slade et al., 2022; Li et al., 2022).

The objective of this research is to develop an end-to-end hardware and software system for ergonomics assessment using custom sensors and free and open software tools, and explore its feasibility in the automotive industry. The main contributions of the present work are: 1) Development of an end-to-end system for ergonomics assessment built with free and open tools; 2) Establishment of its compatibility with a commercial solution throughout different assessment phases; 3) Validation of the proposed system on a workstation in the automotive industry through a comparison of joint angles and a RULA assessment report.

## 2. Methods

### 2.1. Pipeline overview

The pipeline includes both custom hardware and free software for ergonomics assessment of industrial workers. The hardware sensor system can be any of the preferred custom systems that capture absolute orientation data from segments of the human body. The software component of the pipeline is a toolkit, based on open-source and free software frameworks, used for data processing, visualization, and analysis. The toolkit consists of a pyGame or Unity 3D tool for avatar visualization and recording, NumPy, SciPy, and Matplotlib libraries for numerical computation and data visualization, and OpenSim for musculoskeletal modeling and inverse kinematics computing. The toolkit is designed to be user-friendly and extensible, with a modular architecture that allows users to customize and extend its functionality

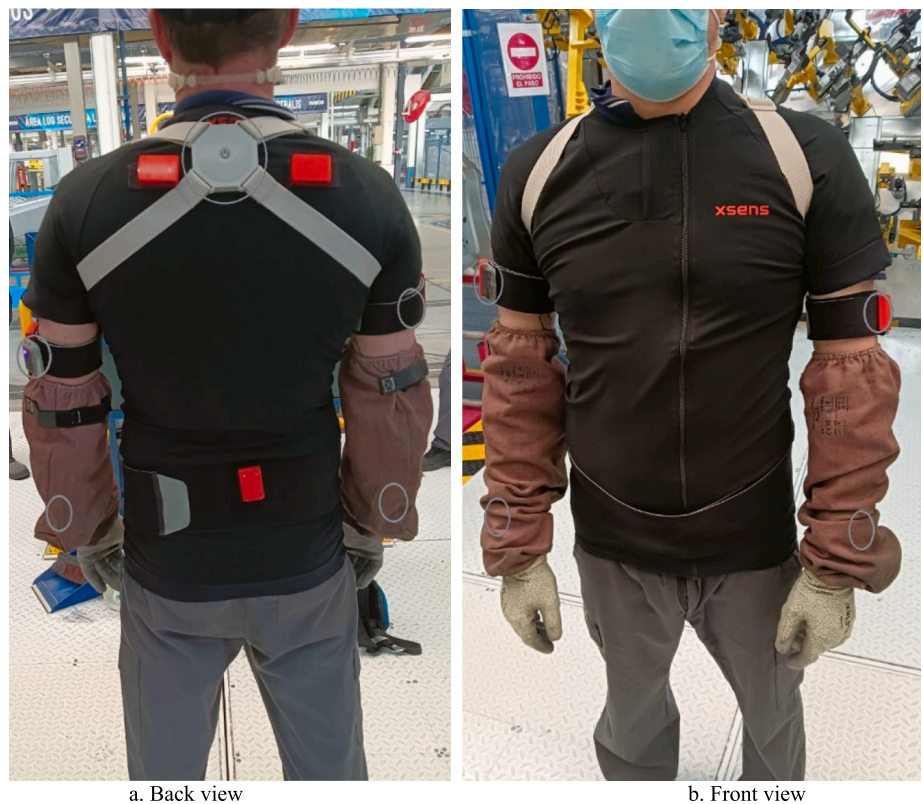

a. Back view                                                                 b. Front view

**Fig. 2.** Simultaneous placement of both sensor systems on the same subject.

according to their specific needs. The pipeline is composed of the modules: Sensors, Movement acquisition, Movement analysis, and Ergonomics analysis (RULA reports), as depicted in Fig. 1. The input to the last two modules can be imported from a commercial solution (e.g., Movella Awinda), for maximum compatibility.

*2.2. Sensors*

The chosen pipeline hardware component is a custom IMU system (González-Alonso et al., 2021), based on low-cost IMU modules and a Nordic Semiconductor nRF52 Series single-board processor. The processing core contains an ARM Cortex M4 with a 2.4 GHz transceiver (7–8 m range). The IMU modules are based on the BNO080 (Laboratories, 2017) chip, which generates quaternion data, representing the orientation of the IMU in 3D space. The rechargeable button cell of the previous model has been replaced with a 120mAh lithium polymer battery, together with a battery charging chip (MCP73831) to increase its autonomy. The low-cost custom sensors employed in the pipeline can record up to 11 body segments and feature protection against electromagnetic interference from industrial ferromagnetic cores and saturated environments in the Industry 4.0 band (2.4 GHz) (González-Alonso et al., 2021). For the experiments in this paper, we used 5 wearable nodes attached to the upper parts of the worker's body, with dimensions of 45 × 28 × 10 mm for each sensor.

This approach can be easily replicated by using any other customized system based on IMUs (He et al., 2022; Slade et al., 2022; Li et al., 2022; Liu et al., 2018; Raghavendra et al., 2017; Costa et al., 2020) which, in any case, would be compatible with the rest of the modules of this solution. In particular, the communication protocol can be modified to suit the specific application environment. Although our model uses a modified implementation of Nordic's proprietary protocol (Enhanced Shockburst) to feature a frequency channel hopping mechanism, other technologies, such as Bluetooth (BT) or Wi-Fi, can be used for sensor-to-sensor communication as long as they achieve an acceptable number of measurements per second for the specific target application.

The IMU chosen for our sensors comes with a proprietary fusion algorithm implemented internally on the BNO080 chip, and has been previously tested in human body motion applications (Stanzani et al., 2020). Alternatively, this algorithm can be replaced by other widely used fusion algorithms such as Extended Kalman Filter (Li and Wang, 2013), Mahony (Mahony et al., 2008), or Madgwick (Madgwick et al., 2011), with other sensors with similar characteristics, such as the LSM9DS1, ICM-20948, MPU9250, among other 9DoF units (He et al., 2022; Li et al., 2022; Liu et al., 2018; Raghavendra et al., 2017).

*2.3. Sensors placement*

Sensor placement is crucial in both upper-limb (Höglund et al., 2021) and lower-limb (Niswander et al., 2020) IMU-based approaches, since incorrect placement can lead to measurement inconsistencies. Therefore, a standardized protocol for sensor placement is essential to obtain a reliable analysis of body movements. For the experiments, a total of 12 IMU sensors were placed on the back, the pelvis, the arms, and forearms to register upper body movements. Fig. 2a and Fig. 2b show commercial (orange) and custom sensors (circles) placement. The worker was recorded simultaneously with 7 commercial Movella Awinda sensors and 5 custom IMU sensors. The commercial sensors were placed according to the manufacturer's guidelines (Technologies, 2016). The custom sensors were placed on the back, tracing an imaginary line that connects the two posterior axillary folds (approximately at the T5–T7 level), and on the middle lateral section of each upper arm, as well as the rear part of each wrist. Sensors were placed on fastening straps, on the specifically designed commercial system's T-shirt, and on a special encapsulation for the back custom sensor. Sensors were positioned under the Personal Protection Equipment (PPE) during recordings. In Fig. 2a and Fig. 2b, they are shown above the PPE for visibility. A more detailed description of the protocol for the custom sensors placement is provided in (Martínez-Zarzuela et al., 2023), where the same custom

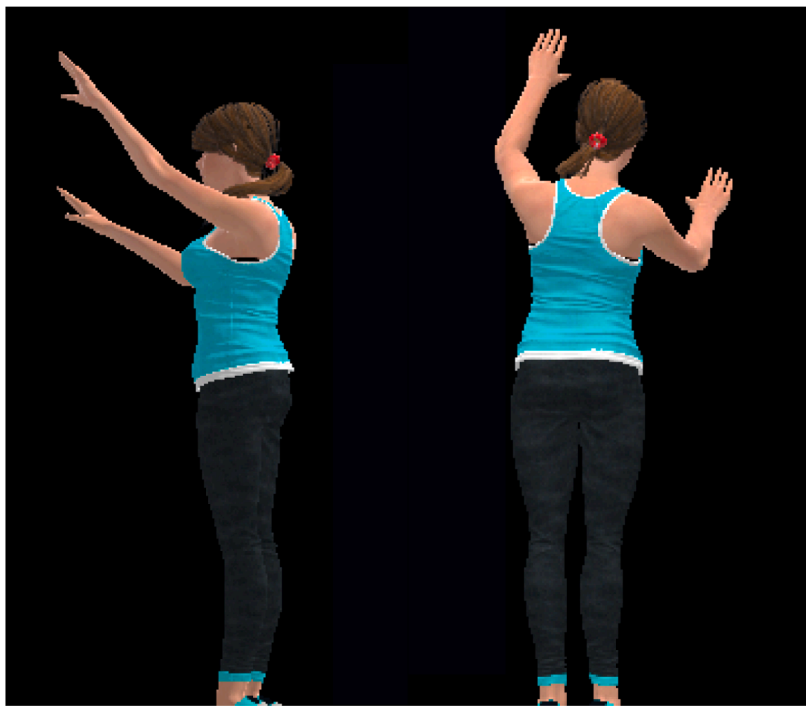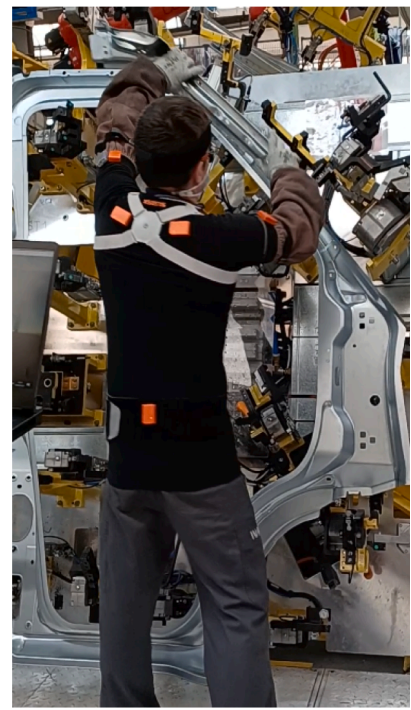

a. First snapshot of the Unity3D visualization and recording tool vs. actual worker pose

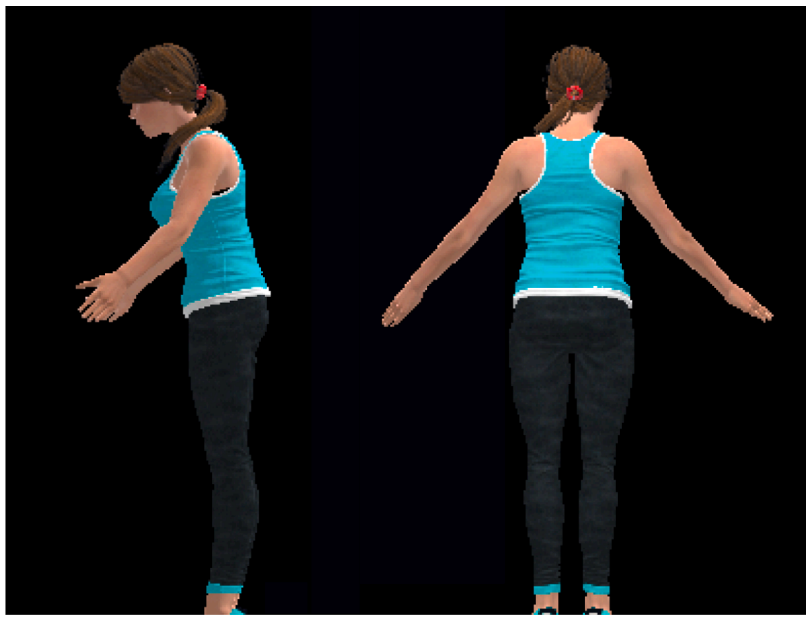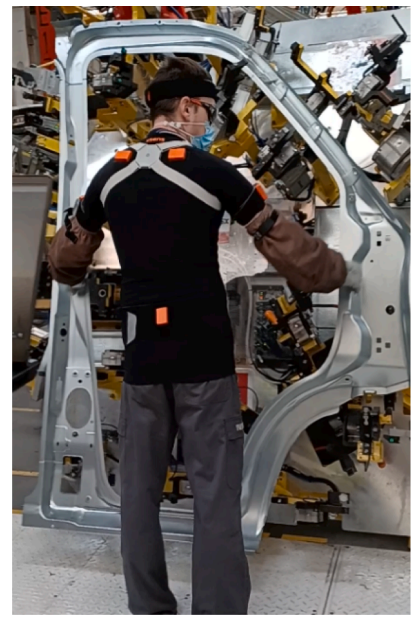

b. Second snapshot of the Unity3D visualization and recording tool vs. actual worker pose

**Fig. 3.** Snapshots of the Unity3D visualization and recording tool (left) and actual worker poses (right).

sensors were used to acquire a database of daily life movements. Commercial and customized sensors were placed at the same approximate locations on the body to obtain comparable results. On the other hand, the commercial solution uses two different sensors on the back and adds an extra sensor on the pelvis. Both systems did not interfere with each other thanks to our band-hopping implementation, which allowed us to obtain comparable data on worker performance, under the same conditions.

According to the literature (Höglund et al., 2021), some planes of motion of certain joints are less affected by sensor placement, including shoulder flexion–extension, abduction–adduction, and elbow flexion–extension. Following the chosen method, only joint range values in certain axes (flexion, abduction, and pronosupination) were considered for the subsequent RULA analysis. Therefore, we will take them as the basis of our study instead of others with a greater influence of the placement, so they will not compute for our subsequent analysis. Due to this, the elbow and shoulder joints were the ones chosen to perform a local ergonomics analysis and their local axis system is defined following the recommendations of the International Society of Biomechanics (ISB) (Wu et al., 2005).

### 2.4. Movement acquisition

The accuracy and reliability of the IMU system and the software pipeline are critical to the validity of the ergonomics assessment results. Therefore, calibration and validation procedures are essential to ensure the quality of the worker's movements and the consistency of the data along the whole pipeline. In the Movements acquisition module, we present a dual-option capture software. The first option includes an avatar representing human joint movements using Unity3D (Fig. 3a and Fig. 3b), and the second a simpler visualization and recording tool of the 3D orientation of each sensor separately using Python, both in real time. These two solutions include recording functions and a calibration of the orientation of the human body bones as described in (González-Alonso et al., 2021), following the IMU-to-Body-Segment alignment guidelines according to (Rajagopal et al., 2016; Maruyama et al., 2020). In this regard, before starting the measurements with the custom solution, we suggest a heading reset of the sensors, and a calibration procedure consisting in the worker adopting an initial pose to start from (commonly N-pose), which is used as a reference for the assumption of the initial body segment orientations. In this software module, the initial orientations of the sensors during calibration are stored in the output files to allow their subsequent use as appropriate inputs of the downstream analysis system. Later on, they are needed for instant pose and kinematics computations as described in (González-Alonso et al., 2021). Most of the outputs of commercial IMU-based solutions, as depicted in Fig. 1, are also compatible with our implementation. As long as the commercial solution provides a quaternion output and a timestamp for each placed sensor, it can be easily adapted to be the input to the Movement analysis module of our pipeline (*Output 1 of Commercial Sensors* in Fig. 1).

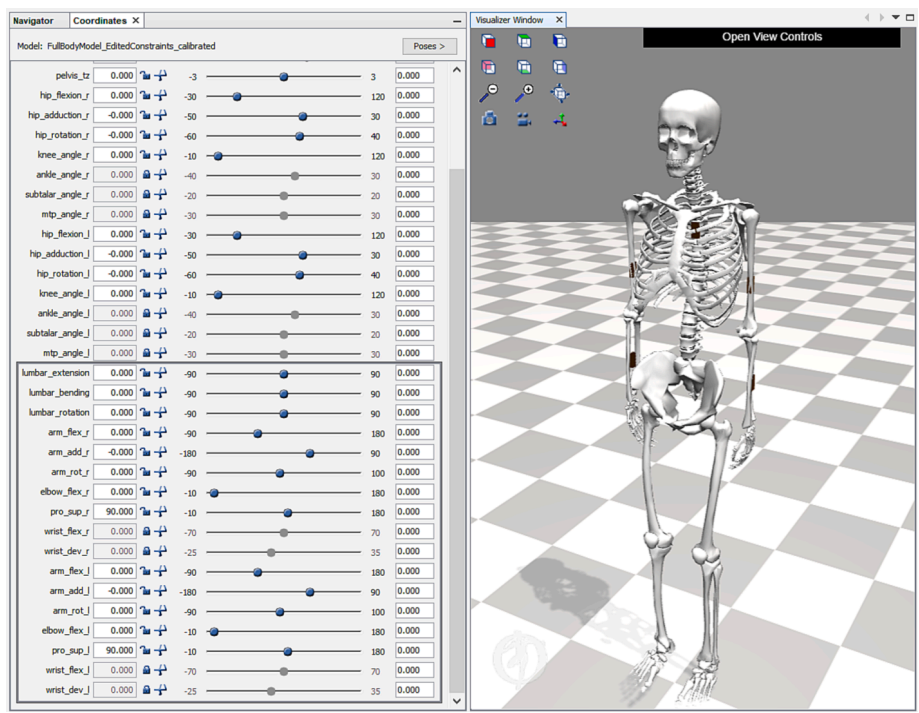

**Fig. 4.** OpenSim model adjusted to N-pose with modified constraints.

At the end of this module, the entire recording is segmented into subsections (frames selection in Fig. 1), and only the frames in which the worker is active are fed into the next module. This is important since inactive moments can introduce unwanted variations in the results of the postural percentage calculation in the final RULA analysis. The capture system described above is used in this step for visual selection of the frames of each subsection by reconstructing the recordings in an avatar.

### 2.5. Movement analysis

The correct implementation of human biomechanical models and inverse kinematics is crucial in human movement research. In the Movement analysis module, joint angle trajectories are inferred from motion capture data from wearable sensors through inverse kinematics. Modeling of body mechanics during complex dynamic movements can be accurately performed with OpenSim, an open-source software platform (Vargas-Valencia et al., 2016) that is validated with different movement capture systems including inertial sensors (Delp et al., 2007). This technology allows the use of biomechanical models for the study of musculoskeletal disorders or rehabilitation procedures, among others.

In our pipeline, previously registered recordings are the inputs to the Movement analysis module, connecting the recorded data to a model of human musculoskeletal structures to generate dynamic simulations of movement. The inverse kinematics output will present the resulting joint angles after applying the constraints of the human model and being adjusted as best as possible to the range of motion achievable by each subject based on the body parameters. In our case, the *Rajagopal2015* model (Al Borno et al., 2022) was modified by adjusting its upper-body constraints to obtain the inverse kinematics through OpenSense (Delp et al., 2007), that has been built on top of OpenSim (Vargas-Valencia et al., 2016). Specifically, wider movements were allowed for the shoulder and arm joints, as the left-side of Fig. 4 highlights, fitting the use-case necessities.

Before ergonomics assessment, the subject is asked to remain motionless and performing an N-pose, as mentioned in the previous section. In Fig. 4, it can be observed that the standard pose of the OpenSim base model (Al Borno et al., 2022) is modified to be the N-pose for the subsequent inverse kinematics process. OpenSim is an open-source software solution that offers endless possibilities for motion analysis. Many research studies have included this software as a basis for their kinematic analysis, and some even have provided clarification on the appropriate procedures to use OpenSim (Mahadas et al., 2019). Alternatively, other inverse kinematics software programs could be used in this module.

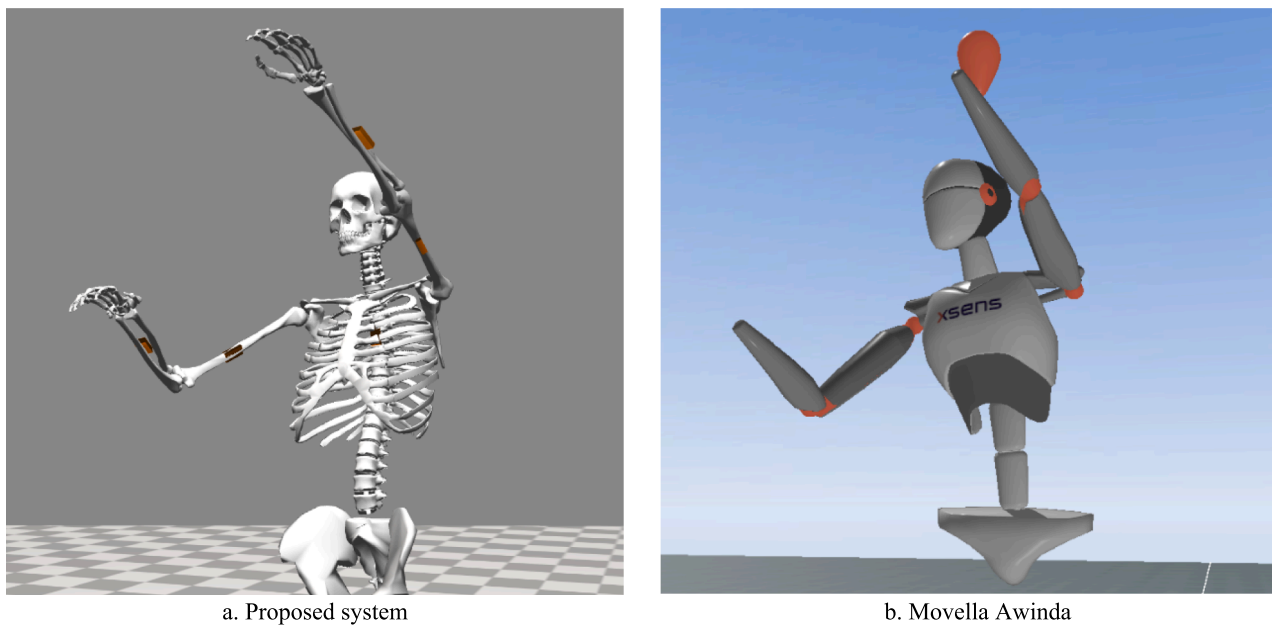

a. Proposed system  b. Movella Awinda

**Fig. 5.** Illustrative avatar view in the same instant for the two systems recorded.

*2.6. Ergonomics analysis and RULA reports*

The Ergonomics analysis module makes it possible to assess the anatomical performance of the worker in a range of activities, highlighting the WMSDs risk of the different postures, which facilitates their subsequent correction. In addition, the software should provide an instantaneous score of the postures produced and a percentage of time spent in each postural risk range with respect to the full work cycle. The output is based on the RULA method (McAtamney et al., 1993) and is designed to facilitate subsequent assessment by ergonomists, streamlining the workstation redesign process and promoting the reduction of occupational accidents.

The Ergonomics analysis module is composed of several Python scripts that assign values to the instantaneous joint angles of the previous module according to the RULA method, providing ergonomics reports in a semi-automatic way. It can be used as the final module for an output file coming from OpenSim, facilitating compatibility with other systems such as optical cameras with markers or Vicon (Panariello et al., 2022) or commercial IMU-based solutions, among others. Commercial systems whose outputs are joint angles can be fed directly into this system module for RULA analysis as shown in Fig. 1 *Output 2 of Commercial Sensors*.

The resulting RULA score corresponds to the values established in the RULA method guidelines for local analysis, considering some adjustments for the final RULA score computation as in (Vignais et al., 2017). The Force/load score was set to 0, considering that the parts handled by the worker were supported by a weightless manipulator (weighing significantly less than 2 kg). In this case, the analysis in subsections does not account for repeatability; therefore, the repeatability score was assigned a value of 0. Since the worker was not raising his shoulders during the task and did not support his arms, these score values were fixed to 0. In addition, it was not considered in the final score if the arm was working across midline or out to the side of the body. Finally, the local joint scores are combined into a global analysis of the total upper body activity. The system calculates the percentage of time (cumulative) that the subject has been within a postural score range. These ranges correspond to risk score levels 1–2, 3–4, 5–6 and 7 and their interpretation follows the guidelines defined by the RULA method (Health and safety statistics, 2021) for the final score. Both left and right sides will be provided, although the worst-case scenario is usually taken to calculate the total postural risk according to the RULA method.

To complete the system, a graphical visualization tool has been developed in Python that displays the instantaneous value of the angles of all the recorded joints, and a traffic light-based color classification system has been introduced for the ergonomics analysis. These results are intended to show at a glance how harmful the movements of a worker are at a given workstation, based on the risk level categories defined by the RULA method. On the one hand, movements grouped in green refer to commonly expected and appropriate movements. On the other hand, orange and red indicate harmful movements that should be performed for the shortest possible time. To show the exact time and percentage of total workstation time for each of these classification levels, a chart was designed as part of the automatic results of the analyzed recording, as it can be seen in Section 3.1. The resulting information is provided to the healthcare professionals in a simplified way so that they can rely on these measurements to deliver a precise and objective ergonomics assessment of the recorded activity.

This way, the ergonomist obtains a first graphical assessment of the local risk score for each joint. We can then interact with the recording using OpenSim, scrolling through the recorded frames until the desired moment is reached. This graphical representation allows ergonomists to explore in detail, for example, the moment of greatest postural risk observing both the model and the RULA score at the same time for comparison. An example is shown in Fig. 5a and Fig. 5b.

**3. Experiments and results**

A case study was conducted to evaluate the pipeline for ergonomics assessment using the RULA method in an automotive assembly line of the IVECO automotive factory in Valladolid, Spain. The analyzed workstation was chosen among the semi-robotized tasks that could be performed by a single operator and involved mainly upper body movements. The workplace involves a wide range of WMSDs risk factors, such as awkward postures and repetitive movements over a long period of time that can be identified by the RULA method. The participant (right-handed male, 36 years old) was a volunteer among the senior workers in that specific part of the assembly line, and signed an informed consent before the experiments. He was asked to perform their work as usual, as part of the daily working hours, while wearing the IMU system on the upper body. The main tasks of this workplace included assembling and disassembling parts, using hand tools, and interacting with robotic arms, and other devices. The activity analyzed corresponds to the one that concentrates the body movements with the greatest injurious load, out of the 2 independent sub-processes into which the factory work line was divided. To determine the time when the recording starts and when it ends, the starting point of each sub-process was taken as the moment when the operator presses the red start button at the workstation and the sub-process ends by pressing the same button to stop. This start-stop process is activated 4 times for each workstation of this factory line. This way we can discriminate the time when the operator is performing work at the workstation as useful time versus rest time or idle time.

**Table 1**
Delta-t values between successive working line subprocesses (min:seconds). Movella Awinda recordings in Frames (recorded at 100 Hz). Custom Sensors data were recorded into two different files (R1 & R2).

| Subprocess | Delta-time Custom Record | Movella Awinda Frames |
| --- | --- | --- |
| 1 | R1 → 0:04–1:04 | F210 – 6300 |
| 2 | R2 → 1:28–2:38 | F24760 – 31,760 |

**Table 2**
Posture summary.

| Subprocess G1 | Left-Elbow | Right-Elbow | Left-Shoulder | Right-Shoulder |
| --- | --- | --- | --- | --- |
| Cross-correlation | 0.958 | 0.971 | 0.961 | 0.955 |
| RMSE | 8.971 | 7.799 | 9.253 | 9.685 |
| Subprocess G2 | Left-Elbow | Right-Elbow | Left-Shoulder | Right-Shoulder |
| Cross-correlation | 0.967 | 0.973 | 0.953 | 0.952 |
| RMSE | 9.580 | 7.412 | 8.667 | 11.48 |

Each task lasted approximately one minute, with pauses in between, and was recorded using the custom IMU system and the Movella Awinda sensors simultaneously. The different tasks were performed and recorded in a row, so the frames of the different tasks captured had to be separated afterwards. The data were processed, visualized, and analyzed using the proposed pipeline, and WMSDs risk factors were assessed using the RULA scoring system.

### 3.1. Joint angles analysis

To evaluate the performance of the approach, the joint angles obtained from the motion captured with our custom system and analyzed with the OpenSim software, were compared to the motion analysis result of the commercial Movella Awinda sensors system with the MVN Analyze Pro license. The measured differences between the joint angles and the RULA outputs are discussed later.

In the capture procedure, the entire recording was supervised by ergonomists from the occupational risk prevention department of IVECO Valladolid. They were asked to pay close attention to the placement of the sensors, and the correct operation of body-worn units during data acquisition in the Unity3D visualization and recording tool. Their participation was crucial to ensure the reliability of the data collected, and therefore the validity of the results and the overall robustness of the study. The registered data outcome consists of quaternion orientations of the worker's body segments, converted into angular variations from the upper body joints through inverse kinematics. The inverse

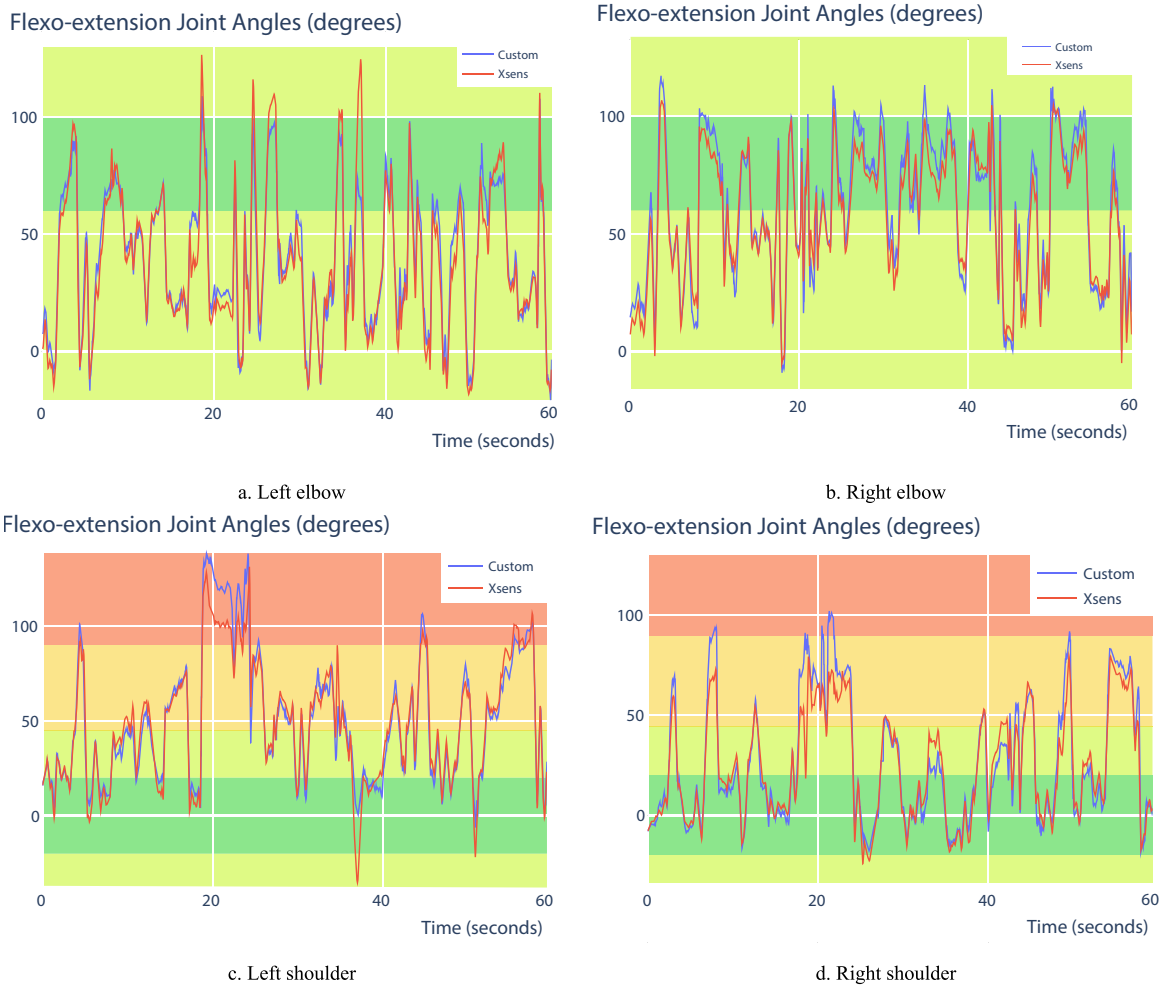

**Fig. 6.** Comparative data for the first subprocess between customized and commercial solutions for the estimated joint angles.

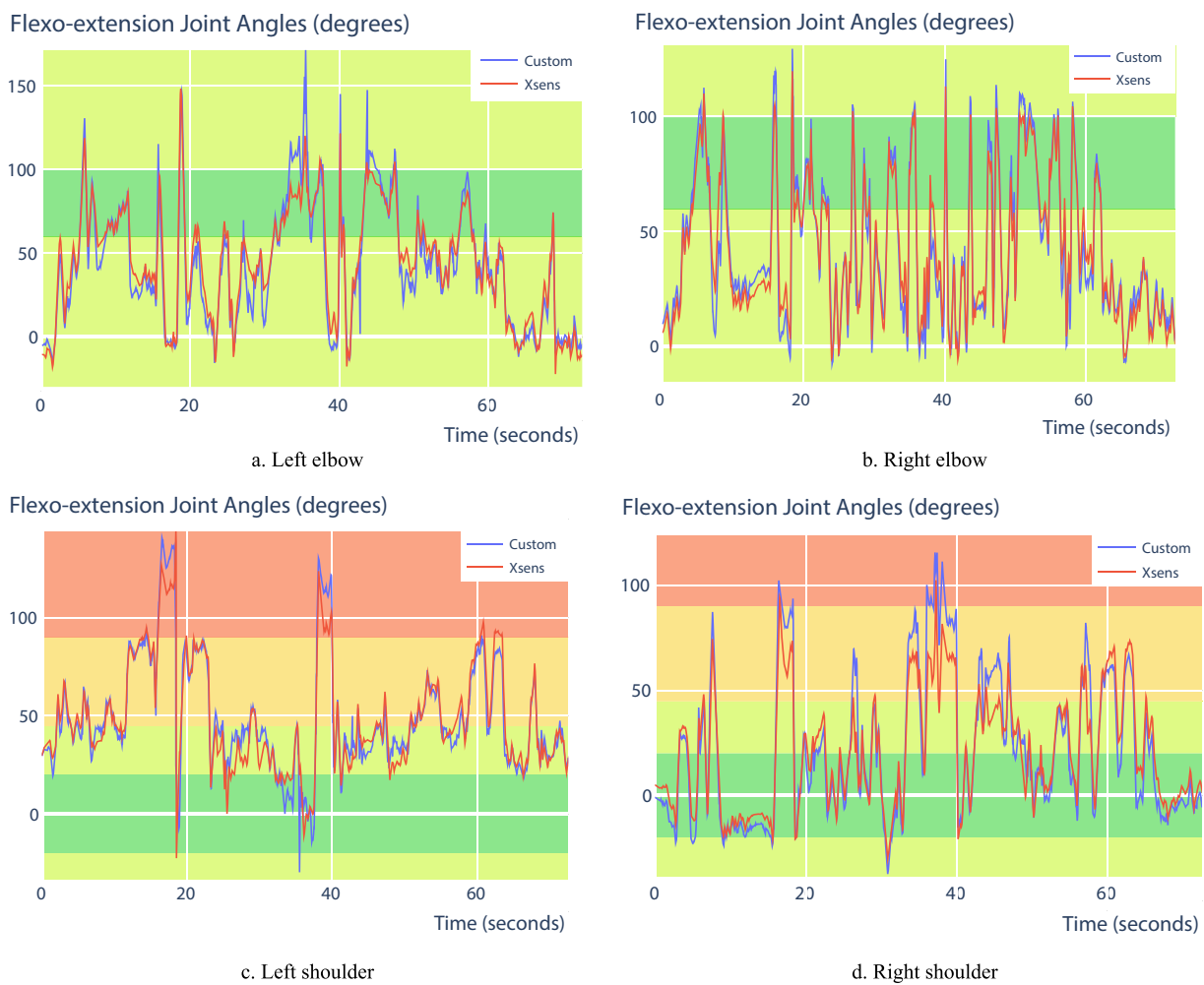

Fig. 7. Comparative data for the second subprocess between customized and commercial solutions for the estimated joint angles.

kinematics process is performed following the process described in Section 2.5.

Records were obtained for the 2 subprocesses of the workstation as defined in Table 1, including flexion–extension and pronation-supination of the elbow, and flexion–extension and abduction of the shoulder, for both the right and left sides of the body.

In a first visual comparative study, we observed how the elbow joint and the shoulder joint on the right and left sides of the body obtained equivalent results. Fig. 5 shows an illustrative capture for the second subprocess. In this experiment, the results in the flexion–extension axis were compared, being this axis the most relevant one used for the subsequent computation of the RULA score.

Table 2 summarizes cross-correlation and Root Mean Square Error (RMSE) results. These parameters are useful for measuring the differences in the joint angles obtained from the proposed pipeline, and those obtained from the gold standard (Output 2 of Commercial Sensors in Fig. 1) for the studied subprocesses.

As it can be seen from the analysis of the resulting inverse kinematics, the motions recorded by both systems (Fig. 6 and Fig. 7) are similar for the elbow joint and only differ slightly in some specific movements. The largest angular differences were found at the shoulder joint (RMSE < 12 degrees) and these differences were not significant for the final RULA score computations.

The RULA score was computed from these angular variations following the guidelines described in Section 2.6. The comparison of results of the two systems considers the fully free pipeline output shown in Fig. 1 versus the MVN Analyze Pro output of joint angles passed through our RULA system, that is, *Output 2 of Commercial Sensors* pipeline.

The final RULA score is presented in two different ways: first, by showing a RULA score at each time instant for each joint, as shown in Fig. 8 for elbow and shoulder flexion, and second, with a risk level percentage of the total cycle time in a pie chart for global results, as shown in Fig. 9 and Fig. 10.

Fig. 8 shows the RULA scores obtained from the joint angles in Fig. 7. It can be observed in Fig. 8 how both systems obtain the same RULA scores, except for very specific instants.

By adding the elbow and shoulder scores, and analyzing each score range over the total time, we derive the overall workstation risk. Thus, based on the established joint angle ranges, we can analyze the overall risk of injury according to the percentages obtained. This analysis is defined as the Global RULA score of a work cycle as shown in Fig. 9 and Fig. 10. As can be seen in both analyses, the two systems obtain similar results for the WMSDs risk assessment at each specific moment and as percentages of the overall risk of the workstation. In addition, a global RULA score table can be obtained for a given observed position of higher risk. The tracking of the achieved position can be ensured by instantaneous representation of the human model in both systems, as shown in Fig. 5a and Fig. 5b.

## 4. Discussion

The conducted study successfully implemented and validated an end-to-end system for ergonomics assessment, including both hardware and software, showing its feasibility in the automotive sector. Other researches (He et al., 2022; Slade et al., 2022; Li et al., 2022) have offered solutions focused only on the hardware, while some others have only focused on the ergonomics assessment (Caputo et al., 2019), but without including a comparison with a reference system. There are also studies that generate RULA assessments using the output of commercial sensors (Huang et al., 2020) or consumer-depth cameras (Manghisi et al., 2017; Abobakr et al., 2019). The presented end-to-end system offers a free and comprehensive ergonomics analysis tool that can indicate a local and global time-based reference score for RULA assessment in the industry. The proposed custom sensors and free tools approach has several attractive features: it constitutes a simple and easy-to-use solution for in situ motion capture; it overcomes the communication limitations of entry-level IMU solutions with respect to crowded 2.4 GHz environments (González-Alonso et al., 2021); it democratizes access to these systems in industry; it provides an adaptation layer between raw sensor data and open software such as OpenSim; and finally, it enables fully customizable solutions to reach the industry by completely adapting to the final workplace.

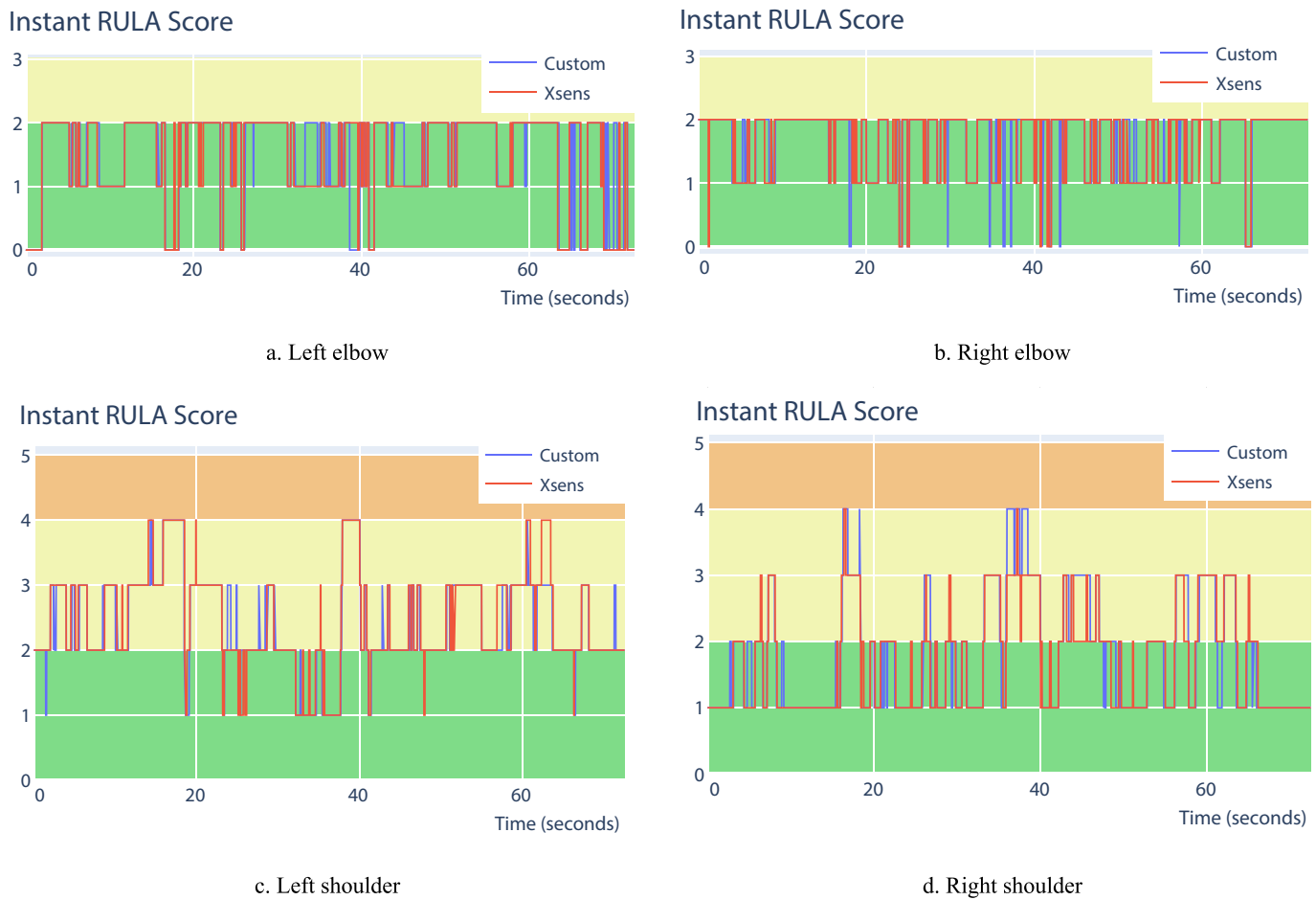

**Fig. 8.** Instantaneous RULA score for the left and right elbows and shoulders in the second subprocess recordings.

The proposed modules through the pipeline can interoperate to some extent with other IMU-based hardware systems and inverse kinematics software solutions, such as the commercial system discussed in this paper. More specifically, Movella Awinda has an export option that provides per-sensor quaternion orientation output (ASCII CSV text format) through MT Manager software. In more advanced commercial solutions, such as the MVN Analyze capture and processing software compatible with the Movella Awinda, a complex calibration process can be performed, which may include a motion routine and obtains a good correction for expected deviations in IMU-based systems. However, the outputs of these systems are proprietary and only compatible with the manufacturer's software (e.g.,.mvn or.mvnx). In MVN Analyze Pro, an export process can be used to export joint angles into an.xlsx file through a license fee. Movella has recently introduced a cloud-based analysis (Movella Motioncloud license) that also incorporates RULA analysis for its recordings, emphasizing the usefulness of these methods and their use with IMU systems.

Custom IMUs and free tools have been successfully used to conduct a semi-automatic ergonomics analysis report and obtain joint angles and RULA scores equivalent to those of a state-of-the-art gold standard solution. Comparative experiments of joint angles computed with the proposed system and the reference system determined that the acquisition and representation of movements in the musculoskeletal model of the pipeline faithfully followed the movements of the recorded subject. The value of the cross-correlation coefficient is higher than 0.95 for elbow and shoulder joint angles and the RMSE is lower than 10 for elbows and 12 for shoulders. We can conclude that the system based on custom wearables and OpenSim obtained inverse kinematics results comparable to those detected by the Movella system. The global RULA score in the different subprocesses analyzed differs by no more than 5 % between both systems' reports.

There are several advantages of using the presented pipeline for ergonomics assessment in the automotive industry. One major strength of this approach is the cost-effectiveness of the pipeline, as it is based on low-cost IMU modules and free developing tools. Although studies involving measurements performed by customized IMU-based solutions are rare in these use cases, this system is presented as an alternative for entry-level IMU systems, which do not perform well in signal-crowded industrial environments (He et al., 2022). Regarding interference-

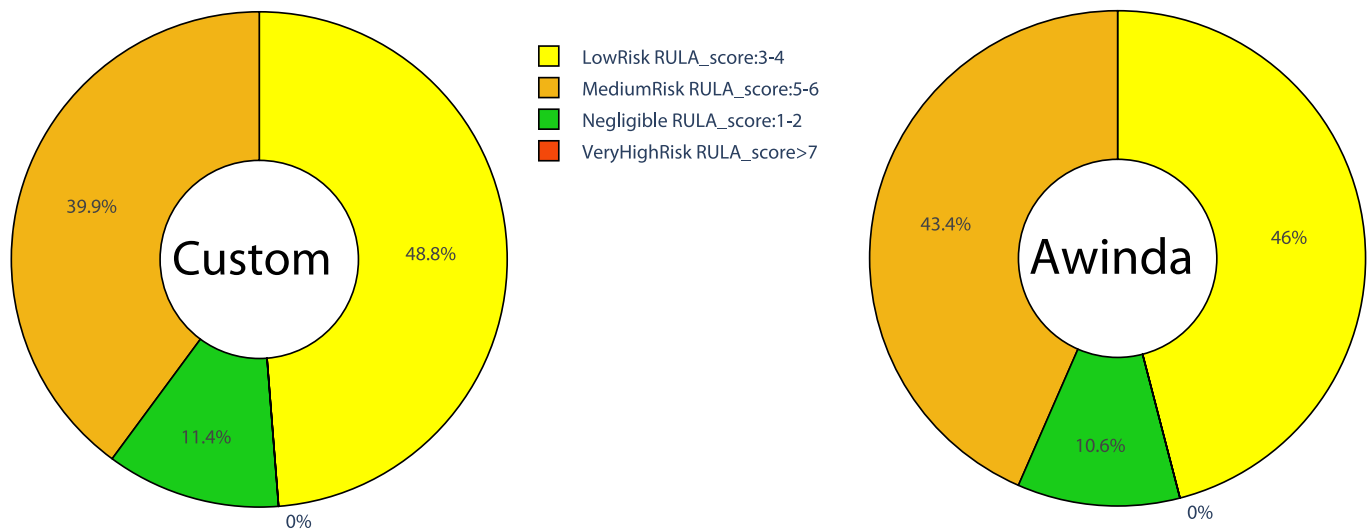

a. Left side of the body.

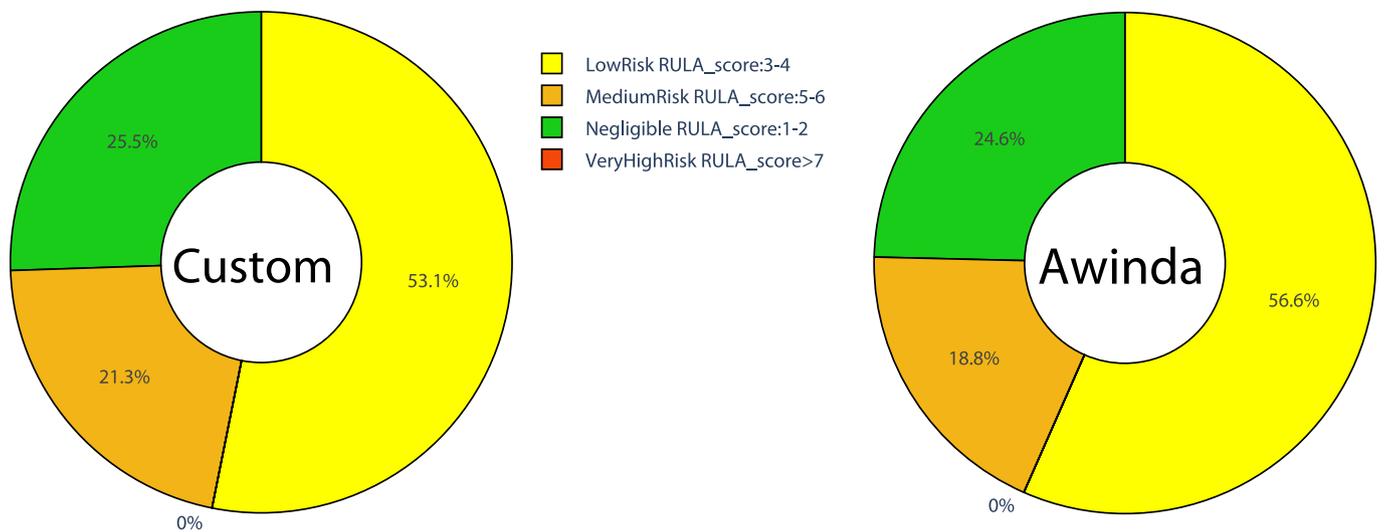

b. Right side of the body.

**Fig. 9.** Comparison of Global RULA scores using the first subprocess recordings obtained through the pipeline with data from the custom sensors and with the output 2 of the commercial sensors (Movella Awinda).

related limitations, in the proposed system the use of a specific communication protocol improves the reliability and safety of the pipeline. In other solutions, interferences causing sensor range reduction and security issues can occur due to the use of standard Bluetooth or other IoT communication technologies. Automotive environments are characterized by high saturation in the 2.4 GHz spectrum band. Besides, the use of free developing tools also enables users to customize and extend the pipeline functionality according to their specific needs and preferences.

One of the main limitations of the experiments conducted is related to the total number of custom IMU sensors employed. Meaningfully, the greater variations observed in the shoulder joint could be mainly due to the placement of a larger number of Movella sensors in this location, specifically the inclusion of one sensor per side on the scapula to obtain data relating to the glenohumeral joint. Future work will add these sensors to the system. To complete the RULA analysis of the upper body, additional sensors can be added. This would allow for a broader study of the trunk and neck, with sensors on the head, chest (already included in this analysis), wrists, and hips, which could also be considered.

Another limitation of the proposed approach is the reliance on the RULA method, which may not be sufficient for more comprehensive and sophisticated ergonomics evaluations. Recent scoring systems, as exemplified by (Ghasemi and Mahdavi, 2020), exhibit superior correlations with WMSDs prevalence compared to traditional methods. These systems provide a more nuanced understanding of risk factors and eliminate sharp angular transitions that cause abrupt changes in the scores of input variables. Although our research did not cover this specific analysis, acknowledging this limitation prompted us to consider

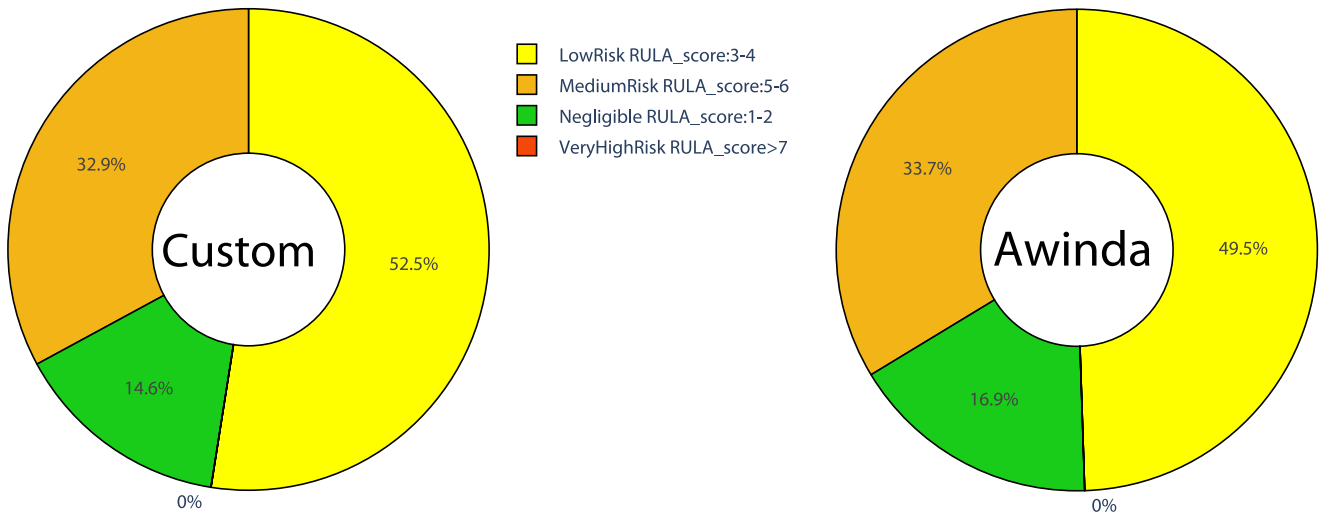

a. Left side of the body.

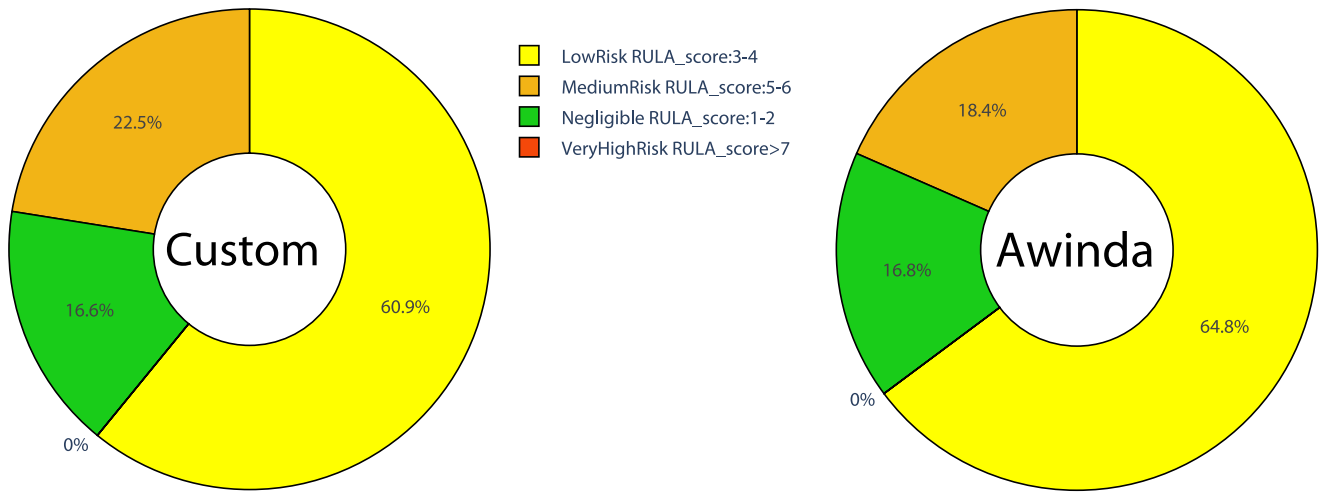

b. Right side of the body.

**Fig. 10.** Comparison of Global RULA scores using the second subprocess recordings obtained through the pipeline with data from the custom sensors and with the output 2 of the commercial sensors (Movella Awinda).

potential enhancements. The adaptability inherent in the modular design of this pipeline allows for such improvements. Consequently, the pipeline could be extended to support other ergonomics assessment methods, such as OWAS (Ovako Working Posture Analyzing System), OCRA (Occupational Repetitive Actions), REBA (Rapid Entire Body Assessment), or EAWS (Ergonomic Assessment Worksheet) or other more novel approaches. Moreover, with the same capture procedure and adding other sensors and tools to the described modules, it would be possible to add more parameters, such as muscle strain or forces involved in musculoskeletal systems and repeatability.

This study is intended to serve as a guide for future research in the use of free tools for the acquisition and analysis of body movements in different application areas. It aims to facilitate and define a path in the different stages of the process that contemplates the possibilities of new open access tools for the study of musculoskeletal disorders through biomechanical models and to develop tools for movement data acquisition and subsequent ergonomics analysis.

In future works we will conduct the recording of a larger number of subjects with different body measurements and positions with different characteristics. More data are necessary to explore the solution in different use cases, as this is a key factor for a generalization of the applicability of the system.

**5. Conclusion**

This study explores the industry applicability of a custom IMU sensor system for measuring human body movements in line with standard ergonomics methods. The proposed IMU-based approach offers easy placement and continuous activity monitoring, enhancing postural assessment and worker re-education. Likewise, human motion data collection from body sensors can aid injury prevention and influence workplace design. In the presented study, the validation of a free ergonomics assessment tool, built on top of low-cost wearable custom sensors, was performed in comparison with a commercial gold standard in an automotive workstation with WMSDs risk. Inverse kinematics analysis showed close agreement between the two models.

The case study demonstrated the feasibility and effectiveness of the pipeline for ergonomics assessment in the automotive industry. The software toolkit was able to provide powerful and flexible tools for data processing, visualization, and analysis, with an average error in the global RULA score of less than 5 % compared to the gold standard IMU system. The proposed pipeline is a promising, cost-effective, and user-friendly tool for assessing ergonomic parameters in the automotive industry, with potential applications in manufacturing, construction, and healthcare. Further research is needed to enhance usability and address any challenges that may arise.

Using free tools and customized systems for recording body movements can popularize objectified, data-driven methods to prevent musculoskeletal injuries, making the presented pipeline a versatile tool for various fields, including preventive medicine, postural training, and musculoskeletal disorder assessment and rehabilitation.

**Informed consent statement**

Informed consent was obtained from the worker involved in the study.

**Funding**

This research has been partially funded by the Department of Employment and Industry of Castilla y León (Spain), under research project ErgoTwyn [INVESTUN/21/VA/0003], by the Ministry of Science and Innovation [PID2021-124515OA-I00], and by a research contract with IVECO Spain SL.

**CRediT authorship contribution statement**

**J. González-Alonso:** Conceptualization, Data curation, Formal analysis, Investigation, Software, Visualization, Writing – original draft. **Cristina Simón-Martínez:** Methodology, Writing – review and editing. **M. Antón-Rodríguez:** Resources, Investigation, Writing – review and editing. **D. González-Ortega:** Resources, Investigation, Writing – review and editing. **F.J. Díaz-Pernas:** Resources, Writing – review and editing. **M. Martínez-Zarzuela:** Conceptualization, Funding acquisition, Investigation, Methodology, Project administration, Software, Supervision, Writing – original draft.

**Declaration of competing interest**

The authors declare that they have no known competing financial interests or personal relationships that could have appeared to influence the work reported in this paper.

**Acknowledgements**

This study was made possible through the funding of the Ministry of Science and Innovation (Spain), the Department of Employment and Industry of Castilla y León (Spain), and a collaboration with the occupational risk prevention department of IVECO Valladolid (Spain). We would like to thank them and IVECO workers for their professionalism during data collection. We would also like to highlight the support given, which allowed us to carry out the work in real industrial workplaces.